\documentclass[prl,twocolumn,preprintnumbers,amsmath,amssymb]{revtex4}
\usepackage{graphicx}
\usepackage{dcolumn}
\usepackage{bm}
\usepackage{mathrsfs}
\usepackage{natbib}
\usepackage{float}

\begin{document}
	
\title{Altering magnetic response of superconductors by rotation}
	
\author{Jun-Ping Wang}
	
\affiliation{Department of Physics, Yantai University, Yantai, P.R. China}
	
\date{\today}
	
\begin{abstract}
		
It is generally believed that, at a certain temperature below the critical one, magnetic response of a superconductor (SC) is determined solely by its intrinsic properties. Here we show that the mechanical rotation of a SC can easily change the values of the critical fields at which the superconductivity is destroyed (type-1 SC) or the vortices penetrate into (exit from) the material (type-2 SC). This is due to a superposition of the Meissner current induced by the external field, and the spontaneous current on the surface of the SC induced by the mechanical rotation. As a result, the critical fields of a SC can be increased or decreased, depending on the geometrical form of the material and the relative orientation of rotation and the external field.

\end{abstract}
	
\maketitle
$Introduction$.
Superconductivity is characterized not only by perfect conductivity, but also by unusual magnetic response \cite{Gennesbook,Tinkham}. In the presence of an external magnetic field, a supercurrent is generated near the surface of the superconductor (SC), which prevents the field from penetrating the interior of it. The external field can destroy the superconductivity as it reaches the thermodynamic critical value $H_{c}$ for a type-1 SC. For a type-2 SC, the magnetic field penetrates into the material in the form of quantum vortices as the field equals to the lower critical value $H_{c1}$, and vortices exit from the SC accompanied by the destruction of superconductivity as the external field increases to the upper critical value $H_{c2}$. 
	
On the other hand, a spontaneous magnetic field can be produced by a rotating SC \cite{Londonbook}. The physical picture is that, as the most of superconducting electrons follow the movement of the rotating body exactly, there are superconducting electrons lag behind near the surface. Those lagged electrons generates a weak current. This current produces a magnetic field, which is homogeneous in the interior of the SC,
\begin{equation}
		\textbf{B}_{L} =-\frac{2mc}{e} {\bf{\Omega}} = \frac{2 c}{\gamma q} \bf{\Omega}. \label{eq: Londonfield}
\end{equation} 
$\textbf{B}_{L}$ is the London field, $m$ and $e$ are the mass and charge of the electron, $c$ is the speed of light, $\mathbf{\Omega}$ is the angular velocity of rotation. Here the absolute value of electric charge $q$ and inverse mass $\gamma$ of two electrons are introduced. The induced current contributes to the effective magnetic moment of the SC - the ``London Moment''. The magnetic field induced by the rotation of the SC has been verified in experiments, both in conventional and high-temperature SCs \cite{Hildebrandt,Brickman,Tate,Verheijen,Sanzari}. Recently, a microscopic dynamic analysis of London moment was performed by Hirsch \cite{Hirsch2013,Hirsch2018a,Hirsch2018b,Hirsch2018c}. 
	
It was shown by Babaev and Svistunov that the rotational response of SCs is different between type-1 and type-2 SCs \cite{babaev2014}. For a type-1 SC, there is a critical rotation frequency at which the SCs experience a first-order phase transition from superconductive to normal state. For a type-2 SC, there exist two critical rotational frequencies at which vortices penetrate into (exit from) the SC. We also note that the vortex lines, which occur in a superfluid enclosed in a cylindrical vessel rotating about its axis, are analogous to vortex lines in rotating type-2 SCs. The number of vortex lines in superfluid increases with increasing rotational frequency. And superfluidity is eliminated above a critical frequency. However, as we show in current study, vortex physics in rotating type-2 SCs in an external magnetic field is not only different from that in stationary type-2 SCs in the magnetic fields, but also different from that in rotating type-2 SCs in the absence of the external field.
	
Here we explore the magnetic response of rotating SCs. It is shown that, the magnetic response of a rotating SC is quite different from that of a stationary one. The effect of superposition of rotation-induced supercurrent (London current) and the current induced by the external field (Meissner current) can change the magnetic response of a SC dramatically. Depending on the the geometrical form of the sample and relative orientation of rotation and the external field, the critical fields of a SC can be increased or decreased. The physical effects predicted here are experimentally observable, especially near the transition temperature $T_{c}$, at which the critical fields are comparable with the London field (\ref{eq: Londonfield}).
	
$A\ simple \ example$.
Let us consider a superconducting sphere which rotates with constant angular velocity  $\mathbf{\Omega}$ in an uniform external magnetic field $\mathbf{H}_{ext}$, and the direction of rotation is parallel to the external field, $\mathbf{H}_{ext} || \mathbf{\Omega}$. To determine the distributions of the supercurrent and the magnetic field, we use the London theory and start with the following equations:
\begin{align}
	\nabla \times \textbf{v} &=\frac{\gamma q}{c} \textbf{h},   \label{eq:London1}\\
		\nabla \times \textbf{h} &=- \frac{4\pi n q}{c} (\mathbf{v-\Omega \times \textbf{r}}).  \label{eq:London2}
\end{align}
Here $\mathbf{v}$ is the velocity of superconducting electrons, $\mathbf{h}$ is the magnetic field, $n$ is the density of the electron pairs. Outside the sphere, $\nabla \cdot \textbf{h} = \nabla \times \textbf{h} =0 $ and $\mathbf{h}(\mathbf{r} \rightarrow \infty)\rightarrow \mathbf{H}_{ext} $. In the case considered here, it is reasonable to assume that the magnetic field $\mathbf{h}$ outside the sphere can be written as
\begin{equation} \label{magout}
		\begin{split}
			h_{r} & =({H_{ext}}+\frac{2M}{r^3})\cos{\theta},\\
			h_{\theta} & =(-{H_{ext}}+\frac{M}{r^3})\sin{\theta}, \\
			h_{\phi} & =0.
		\end{split}
\end{equation}
Here $M$ is a constant, and can be regarded as the induced magnetic moment of the sphere. Inside the sphere, the equation for velocity of the superconducting electrons $\mathbf{v}$ can be deduced from (\ref{eq:London1}) and (\ref{eq:London2}):
\begin{align}
	\nabla \times \nabla \times (\mathbf{v}- \mathbf{\Omega} \times \mathbf{r}) = -{\beta}^{2} (\mathbf{v} -\mathbf{\Omega} \times \mathbf{r}). \label{eq: velocity}
\end{align}
Here ${\beta}^{2} = 4 \pi n \gamma q^{2}/c^{2} $, and ${\beta}^{-1}$ is the London penetration depth. We have used the relations $\nabla \times (\mathbf{\Omega} \times \mathbf{r}) =2 \mathbf{\Omega},  \nabla \times \nabla \times (\mathbf{\Omega} \times \mathbf{r}) = 0 $. Assuming that the velocity of the superconducting electrons has only the azimuthal component $\mathbf{v}=v \mathbf{e_{\phi}}$, the equation (\ref{eq: velocity}) can be solved:
\begin{equation}
		\mathbf{v} = \left[   \Omega r + \frac{A}{r^2}(\sinh\beta r - \beta r  \cosh\beta r) \right]   \sin\theta  \mathbf{e_{\phi}}. \label{eq: vphi}
\end{equation}
$A$ is a constant to be determined. Substituting equation (\ref{eq: vphi}) into equation (\ref{eq:London1}), the magnetic field inside the sphere is obtained,
\begin{equation} \label{maginside}
		\begin{aligned}
			h_{r} & = \frac{c}{\gamma q} \left[   2\Omega + \frac{2A}{r^3}(\sinh\beta r - \beta r  \cosh\beta r) \right] \cos \theta,  \\
			h_{\theta} & = \frac{c}{\gamma q} \left[  - 2\Omega + \frac{A}{r^3}( (1 + \beta^{2} r^{2})\sinh\beta r - \beta r  \cosh\beta r) \right] \sin \theta, \\
			h_{\phi} & = 0.
		\end{aligned}
\end{equation}
With the continuity of magnetic field at the boundary of the sphere, the constants $M$ and $A$ can be determined,
\begin{align}
		A & = -\frac{3}{2} (H_{ext}-B_{L}) \frac{\gamma q R}{ c \beta^{2} \sinh \beta R},  \label{eq:A}\\
		M & = -\frac{R^3}{2} (H_{ext}- B_{L} )\left[1+ \frac{3}{\beta^2 R^{2}} (1- \beta R \coth \beta R) \right]. \label{M}
\end{align}
$B_{L}=2 c \Omega /{\gamma q}$, $R$ is the radius of the sphere. Thus the distributions of the magnetic field $\mathbf{h}$ and the velocity field ${\mathbf{v}} $ of the superconducting electrons  in all spaces are given.
	
\begin{figure}[t] 
		
	\includegraphics[width=0.53\textwidth, height=7.4cm]{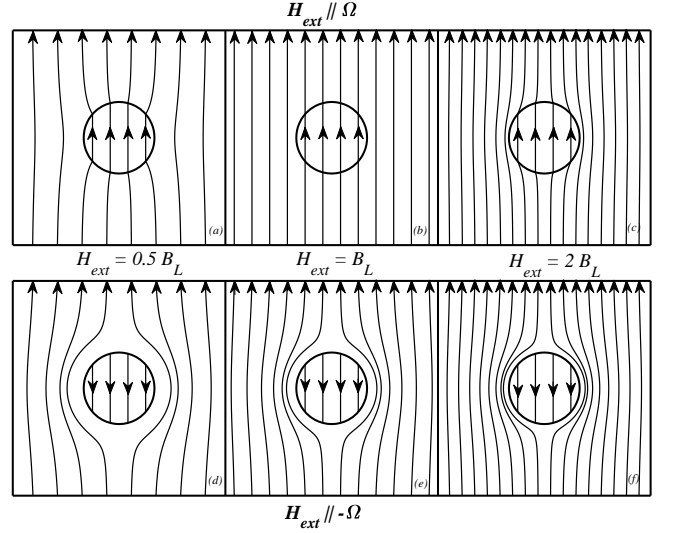}
		
\caption{\label{fig:sphere} A rotating superconducting sphere in the external field. Except for a layer of the London depth near the surface, the magnetic field inside the sphere is uniform $\mathbf{B}_{L}=2 c \mathbf{\Omega}/ {\gamma} q$ (London field). The magnetic field outside the sphere is the superposition of the external field and the field generated by the induced magnetic moment $M$ of the sphere. The induced magnetic moment $M$ is the result of the combined effect of the external field and the rotation. From left to right: the external field increases from $H_{ext}=0.5 B_{L}, B_{L} \ {\text{to}} \ 2 B_{L}$. The top row shows the cases that the field is parallel to the angular velocity. (a) Magnetic field lines converge toward and penetrate through the sphere. (b) The Meissner current is eliminated by the London current and the total supercurrent vanishes. The magnetic field is uniform in the whole space. (c) Magnetic field lines tend to spread out around the sphere. The bottom row shows the cases that the field is antiparallel to the angular velocity. In all three cases, (d)-(f), magnetic field lines are repelled by the sphere strongly comparing to the stationary cases. Note that the direction of the field reverses at the equator of the sphere. Parameters used: London penetration depth ${\beta}^{-1}=10^{-5}cm$, radius of the sphere $R =1 \ {cm}$, angular velocity $\Omega=10^{3} \ sec^{-1}$, London field $B_{L}=1.14 \times 10^{-4} \ Gauss$. }

\end{figure}
	
The supercurrent ${\mathbf{J}}$ can be extracted from (\ref{eq: vphi}) and (\ref{eq:A}) by taking the rotation of the sphere into consideration, $\mathbf{J}=-n q (\mathbf{v-\Omega \times \textbf{r}})$. The current is confined near the surface of the SC with a depth $~\beta^{-1}$, and equals to the sum of the Meissner current ${\mathbf{J}_{M}}$ induced by the external field $\mathbf{H}_{ext}$, and the London current ${\mathbf{J}_{L}}$ induced by the mechanical rotation of the sphere with angular velocity $\mathbf{\Omega}$:
\begin{gather}
		\mathbf{J} =\mathbf{J}_{M} + \mathbf{J}_{L},  \label{eq:totalc} \\
		\mathbf{J}_{M} = -n q \frac{{A}_{1}}{r^2}(\sinh\beta r  - \beta r  \cosh\beta r)    \sin\theta  \mathbf{e_{\phi}}, \label{eq:Mc} \\
		\mathbf{J}_{L} =- n q \frac{{A}_{2}}{r^2}(\sinh\beta r - \beta r  \cosh\beta r)    \sin\theta  \mathbf{e_{\phi}}. \label{eq:Lc} 
\end{gather}
Here
\begin{align}
		A_{1} & = -\frac{3}{2} H_{ext} \frac{\gamma q R}{ c \beta^{2} \sinh \beta R},  \label{eq:A1}\\	
		A_{2} & =\frac{3}{2} B_{L} \frac{\gamma q R}{c \beta^{2} \sinh \beta R}.  \label{eq:A2}
\end{align}
Note that the directions of these two current are opposite. Except for a layer of the depth $\beta^{-1}$ near the surface of the sphere, the magnetic field inside the sphere (\ref{maginside}) equals to the field induced by the rotation of the superconducting sphere with angular velocity $\mathbf{\Omega}$. The magnetic field outside the sphere (\ref{magout}) is the superposition of external field $\mathbf{H}_{ext}$ and the field generated by the induced magnetic moment $M$ of the sphere. It is clear from (\ref{M}) that this induced magnetic moment of the sphere is the result of the combined effect of external field and the rotation: 
\begin{gather}
		M = M_{M} + M_{L},	\\
		M_{M} = -\frac{R^3}{2} H_{ext}\left[1+ \frac{3}{\beta^2 R^{2}} (1- \beta R \coth \beta R) \right],\\
		M_{L} = \frac{R^3}{2} B_{L}\left[1+ \frac{3}{\beta^2 R^{2}} (1- \beta R \coth \beta R) \right].
\end{gather}
$M$ is the total moment, $M_{M}$, $M_{L}$ are the magnetic moments generated by the Meissner current $\mathbf{J}_{M}$ and the London current $\mathbf{J}_{L}$ respectively. In figure (\ref{fig:sphere}) we illustrate the distributions of the magnetic field for a rotating superconducting sphere in the external field.  
	
The effect of mechanical rotation of the sphere on its magnetic response originates from the London current $\mathbf{J}_{L}$. It can be verified from (\ref{eq:Lc}) and (\ref{eq:A2}) that the London current $\mathbf{J}_{L}$ equals to the current generated by a fictitious external field $\mathbf{H}_{fic}=-\mathbf{B}_{L}= - {2 c} {\bf{\Omega}}/{\gamma} q $. Except for a layer of the penetration depth $\beta^{-1}$ near the surface, the magnetic field generated by the London current inside the sphere, $\mathbf{B}_{L}$, is uniform. Outside the sphere, the London current generates a magnetic field equals to that of a sphere with London moment $M_{L}$.      
	
Due to the occurrence of the London current $\mathbf{J}_{L}$, one may speculate that a rotating superconducting sphere with angular velocity $\mathbf{\Omega}$ in the external field $\mathbf{H}_{ext}$ is equivalent to a stationary sample in the external field $\mathbf{H}_{ext}+\mathbf{H}_{fic}$. Here $\mathbf{H}_{fic} = -\mathbf{B}_{L} $ is the fictitious field to produce the London current $\mathbf{J}_{L}$. We will show in the next section that this is indeed the case if the demagnetization factor $N$ of the sample is zero. For a type-1 superconducting sphere with $N=1/3$, it is equivalent to a stationary sample in the external field $\mathbf{H}^{'}$ and 
	
\begin{equation}
		\begin{split}
			\int d \mathbf{r} & (g_{s}(\mathbf{H}_{ext}, \mathbf{\Omega})-g_{n}(\mathbf{H}_{ext}))=\\
			& \int d \mathbf{r}  (g_{s}(\mathbf{H}^{'})-g_{n}(\mathbf{H}^{'})).
			\label{eq: gibbs s} \\
		\end{split} 
\end{equation}
$g_{s}$ and $g_{n}$ are Gibbs energy densities of the superconducting state and normal state respectively. Be aware of the perfect diamagnetism in a stationary sample, it can be proved easily that 
	
\begin{equation}
		H^{'} = \sqrt{H^{2}_{ext}+2B^{2}_{L}-3H_{ext}B_{L}}. \label{eq: Hbar}
\end{equation}
In most cases $H^{'} < H_{ext}$, since $B_{L}$ is small and $H_{ext}>2B_{L}/3$. In the above example, the critical field for a type-1 superconducting sphere changes as $H^{'}=H_{c}$. As the external field equals to  $H_{ext}=2(H_{c}+B_{L}/2)/3$, it can be verified from (\ref{magout}) and (\ref{maginside}) that the field at the equator of the sphere equals to the critical field $H_{c}$. This means that the small part of the sphere of a type-1 SC near the equator tends to be in the normal state. As a comparison, the field near the equator of the sphere reaches the critical value $H_{c}$ when the external field increases to $2H_{c}/3$ in the stationary case. The rotation of the superconducting sphere does change its critical field in this example.

$General \ consideration$.
To study general aspects of the magnetic response of a rotating SC, we consider the  minimal model of the superconductivity in the rotating frame
\begin{equation}
		f_{s} = \frac{\gamma n \hbar^{2}}{2} \left[\nabla \theta + \frac{q}{\hbar c} (\mathbf{A} - \frac{c}{\gamma q} \mathbf{\Omega} \times \mathbf{r}) \right]^{2} -\frac{{\mathbf{H}}_{c}^{2}}{8 \pi} + \frac{1}{8 \pi}(\nabla \times \mathbf{A})^2. \label{eq: free energy}
\end{equation} 
$f_{s}$ is the free energy density, $\theta$ is the phase of the order parameter, $\mathbf{A}$ is the vector potential. Here we take the density $n$ as a constant and omit the free energy density of the normal state in the absence of the magnetic field. The first term in equation (\ref{eq: free energy}) is the kinetic energy, the second term is the condensation energy, and the field energy is presented in the last term. The free energy density (\ref{eq: free energy}) can be rewritten as follows:
\begin{equation}
		f_{s} = \frac{\gamma n \hbar^{2}}{2} (\nabla \theta + \frac{q}{\hbar c} {\mathbf{A}}_{eff} )^{2} -\frac{{\mathbf{H}}_{c}^{2}}{8 \pi} + \frac{1}{8 \pi} [(-\mathbf{B}_{L})-\nabla \times {\mathbf{A}}_{eff}]^2. \label{eq: free energy2}
\end{equation}
Here $\mathbf{B}_{L} = c \nabla \times (\mathbf{\Omega} \times \mathbf{r})/ \gamma q = 2 c \mathbf{\Omega} / \gamma q$ is the London field,
\begin{equation}
	{\mathbf{A}}_{eff} = \mathbf{A} - \frac{c}{\gamma q} \mathbf{\Omega} \times \mathbf{r} \label{eq: A eff}
\end{equation}
is an effective vector potential. The energy density (\ref{eq: free energy2}) is equal to the Gibbs energy density difference for a SC between the superconducting state and the normal state, both in the external field $\mathbf{H}_{fic}=-\mathbf{B}_{L}$. Based on this equivalence, the rotating SC (\ref{eq: free energy}) and (\ref{eq: free energy2}) can be regarded as a stationary SC under the fictitious field $\mathbf{H}_{fic}$, and some insightful results have been obtained in \cite{babaev2014}.   
	
\begin{table*}[ht]
		\caption{The comparison of supercurrents, distributions of the field and critical fields in three cases} \label{table:table1}
		\begin{tabular}{p{4.5cm}  p{4.3cm} p{3.7cm} p{5.0cm}}
			\hline \hline
			& SC in an external field  & Rotating SC &Rotating SC in an external field \\
			\hline 
			Variables & $\mathbf{H}_{ext}$ & $\mathbf{\Omega}$ & $\mathbf{H}_{ext}, \ \mathbf{\Omega}$ \\
			Supercurrents & $\mathbf{J}_{M}$ & $\mathbf{J}_{L}$ &  $\mathbf{J}_{M} + \mathbf{J}_{L}$ \\
			Magnetic field inside the SC & $\mathbf{B}=0$  & $\mathbf{B}_{L}= 2 c \mathbf{\Omega}/ \gamma q$ & $\mathbf{B}_{L}= 2 c \mathbf{\Omega}/ \gamma q$ \\
			Magnetic field outside the SC & $\mathbf{H}_{ext} +$ field generated by $\mathbf{J}_{M}$ & field generated by $\mathbf{J}_{L}$ & $\mathbf{H}_{ext} +$ field generated by $\mathbf{J}_{M}$, $\mathbf{J}_{L}$  \\
			Critical field for type-1 SC & $H_{c}=H_{c}(T)$ & $H_{c}=H_{c}(\Omega,T)$ & $H_{c}=H_{c}(T) \pm 2 c \Omega / \gamma q$ \\
			Critical fields for type-2 SC & $H_{ci} (i=1,2)=H_{ci}(T) $ & $H_{ci} (i=1,2)=H_{ci}(\Omega, T) $ & $H_{ci} (i=1,2)=H_{ci}(T) \pm 2 c \Omega / \gamma q $ \\
			\hline \hline
		\end{tabular}	
\end{table*}

Taking into account the external field $\mathbf{H}_{ext}$, the Gibbs energy density of a rotating SC is \cite{wang}
\begin{equation}
		g_{s}=f_{s}-\frac{1}{4 \pi} \mathbf{H}_{ext} \cdot [(\nabla \times \mathbf{A}_{eff})+ \mathbf{B}_{L} ]. \label{eq: gibbs 1}
\end{equation} 
The demagnetization factor here is set to zero, $N=0$. It can be realized in, for example, a long cylinder in an axially applied magnetic field. Then $\mathbf{H}$ is uniform in all spaces. The Gibbs energy density of the normal state under the external field $\mathbf{H}_{ext}$ is 
\begin{equation}
		\begin{split}
			g_{n}  = \frac{{\mathbf{H}}^{2}_{ext} }{8 \pi} - \frac{1}{4 \pi} \mathbf{H}_{ext} \cdot {\mathbf{H}}_{ext}= - \frac{{\mathbf{H}}^{2}_{ext} }{8 \pi}.
			\label{eq: gibbs n}
		\end{split}
\end{equation}    
The Gibbs energy density difference of the rotating SC between the superconducting state and the normal state, both in the external field $\mathbf{H}_{ext}$, can be obtained from (\ref{eq: free energy2}), (\ref{eq: gibbs 1}) and (\ref{eq: gibbs n}): 
\begin{equation}
		\begin{split}
			g_{s}&-g_{n} = \frac{\gamma n \hbar^{2}}{2} (\nabla \theta + \frac{q}{\hbar c} {\mathbf{A}}_{eff} )^{2} -\frac{{\mathbf{H}}_{c}^{2}}{8 \pi}  \\
			& + \frac{1}{8 \pi} [ (\mathbf{H}_{ext}+(-\mathbf{B}_{L})) - (\nabla \times \mathbf{A}_{eff}) ]^{2}. \label{eq: gibbs d}
		\end{split}
\end{equation}
The equations which determine the motion of supercurrent and the distribution of the magnetic field can now be obtained by minimizing the energy difference,
\begin{equation}
		\delta \int d \mathbf{r} (g_{s}-g_{n}) =0. \label{delta}
\end{equation}   
It is clear from (\ref{eq: gibbs d}) and (\ref{delta}) that a rotating superconducting SC with angular velocity $\mathbf{\Omega}$ in the external field $\mathbf{H}_{ext}$, and a stationary SC in the external field $\mathbf{H}_{ext}+\mathbf{H}_{fic}= \mathbf{H}_{ext}-\mathbf{B}_{L}$, are physically equivalent.

Equations (\ref{eq: gibbs d}) and (\ref{delta}) describe the underlying physics of a rotating SC in the external field. The effect of rotation of the SC on its magnetic response is reflected in the fictitious field $\mathbf{H}_{fic}=-\mathbf{B}_{L}$. The critical fields, at which superconducting-to-normal phase occurs in a type-1 SC, or the vortex phase transitions occur in a type-2 SC, can be increased or decreased, depending on the relative orientation of the external field and the angular velocity of the rotation of the SC: 
\begin{align}
		\mathbf{H}_{c} & \rightarrow \mathbf{H}_{c}-	\mathbf{H}_{fic} = \mathbf{H}_{c} +	\mathbf{B}_{L}, \label{eq: type1} \\
		\mathbf{H}_{ci} & \rightarrow  \mathbf{H}_{ci} -	\mathbf{H}_{fic}  = \mathbf{H}_{ci} +	\mathbf{B}_{L}, (i=1,2). \label{eq: type2}
\end{align}     
	
Vortex physics in a rotating type-2 SC under external field is different from that in a stationary one. If the external field is parallel to the angular velocity of the rotation, both of the lower critical field and the upper critical field are increased the amount by $B_{L}$. As the external field equals to $H_{c1}+B_{L}$, vortices start to penetrate into the SC. Since there exists the London field $\mathbf{B}_{L}$ inside the SC, the flux through a vortex is the sum of the quantum flux $\Phi_{0}=h c/q$ and the flux from the London field, ${\Phi}={\Phi_{0}}+{B_{L}}\Delta S$, where $h$ is the Planck constant, $\Delta S$ is the effective area of a vortex. With increasing field, vortex lattice form and distance between vortices decreases. The vortices exit from the SC as the external field increases to $H_{c2}+B_{L}$.
	
If the external field is antiparallel to the angular velocity of the rotation, both the lower and the upper critical field decreased the amount by $B_{L}$. As the external field equals to $H_{c1}-B_{L}$, vortices penetrate into the SC. The quantum flux carried by vortices here is $antiparallel$ to the London field inside the SC. As a consequence, the total flux inside the SC is decreased the amount by the integer multiples of quantum flux $\Phi_{0}$. With increasing field, invasion of the magnetic field in the form of vortices will eliminate the London field and there is no flux inside the SC. As the external field increases further, magnetic flux inside the SC reverse and parallel to the external field. Vortices exit from the SC as the external field equals to $H_{c2}-B_{L}$.
	
Equations (\ref{eq: free energy}) and (\ref{eq: gibbs d}) are derived in the rotating frame. And it can be verified that the start equations (\ref{eq:London2}) and (\ref{eq: velocity}), which are used to study the rotating superconducting sphere in the previous section, can be deduced from these equations. The choice of the reference frame does not change the conclusions about the magnetic response of a SC. Of course, all physical observable quantities should be translated into the experimental stationary frame. In the present case, the physical picture in the experimental stationary frame and that in the rotating frame fixed on the SC is different in the supercurrent velocity and possible global rotation of the vortex lattice in a type-2 SC.

$Discussion \ and \ Conclusion$.
London field generated by the mechanical rotation of the SC is generally weak. For the angular velocity $\Omega \approx 10^{3} \ sec^{-1}$, the London field is of the order of $10^{-4} \ Gauss$. To improve the observability of the effects predicted in the present work, the experiments can be performed near the transition temperature $T_{c}$. Since the critical fields of the SCs decreases with the increasing  temperature, e.g., $H_{c}(T) \propto  1-(T/T_{c})^{2} $ for type-1 SCs, it can be expected that the London field becomes comparable with the critical fields $H_{c}$, or $H_{ci} \ (i=1,2)$ near the transition temperature $T_{c}$. Then the effects predicted in this work could be significant and observable.         
	
Here we concentrate on the combined effects of external magnetic field and mechanical rotation on the SCs. If we set the angular velocity of the rotation to zero, $\Omega \rightarrow 0$, the classical magnetic response of SCs is recovered. On the other hand, in the absence of the external field, $H_{ext}=0$, the rotational response of the SCs is also nontrivial \cite{babaev2014}. As we have seen, magnetic response of rotating SCs strongly depend not only on the geometrical form of the SC samples but also on the relative orientation of rotation and the external field. In general cases, one should perform detailed calculations to analyze the magnetic response of the rotating SCs. In the simplest case, in which a long cylinder sample placed in an axially applied magnetic field and the demagnetization effect does not exist, a table (\ref{table:table1}) is presented to compare physical quantities in three cases. 
	
Finally, the validity of the free energy expression in the rotating frame ($\ref{eq: free energy}$) should be addressed. In (\ref{eq: free energy}), only the relative velocity difference between superconducting electrons and the crystal lattice of positively charged ions in the rotating frame is taken into account. In fact, the influence of fictitious forces in the rotating frame on the system should also be considered. Here we discuss the possible influence of fictitious forces in an axisymmetric SC, in which the velocity of the superconducting electrons has only the azimuthal component. There are three kinds of fictitious forces. Two of them, the Coriolis force $\propto 2 \mathbf{v^{'}} \times \mathbf{\Omega} $ and the centrifugal force $ \propto -\mathbf{\Omega} \times \mathbf{\Omega} \times \mathbf{r}$, contribute nothing to the energy of the superconducting electrons. $\mathbf{v^{'}}$ is the velocity of superconducting electrons in the rotating frame. The third force, the Euler force $\propto \mathbf{r} \times d \mathbf{\Omega} / d t$ can accelerate or decelerate the superconducting electrons, depending on the angular acceleration of the rotation, and does contribute to the energy (\ref{eq: free energy}). To guarantee the validity of the free energy (\ref{eq: free energy}), one may adjust the angular velocity of the rotating SC slowly to ensure $d \mathbf{\Omega} / d t \rightarrow 0$. 
	
In conclusion, we investigate the magnetic response of the SCs under rotation. Depending on the geometrical form of the SCs and relative orientation of rotation and the external field, critical fields of the SCs, at which the superconducting-to-normal phase transition (type-1 SC) or the vortex phase transitions (type-2 SC) occur,  can be increased or decreased. The results in the present work show that mechanical rotation influences the phase transitions driven by the magnetic field in the SCs.
	
$Acknowledgements$.
This work was supported by the Shandong Provincial Natural
Science Foundation, China (Grant No. ZR2011AQ025) and the National
Natural Science Foundation of China (Grant No. 11104238).

\end{document}